\documentclass[a4paper,fleqn]{cas-dc}

\usepackage[numbers]{natbib}
\usepackage{graphicx}
\usepackage{booktabs}
\usepackage{amsmath}
\newcommand{\rmspace}{\vspace{-2ex}}

\usepackage{tabularx}
\usepackage{xspace,mfirstuc,tabulary}
\usepackage{enumitem}
\usepackage{pifont}
\newcommand{\cmark}{\ding{51}} % check mark
\newcommand{\xmark}{\ding{55}} % cross mark
\usepackage{multicol}
\usepackage{lastpage}

\begin{document}
\let\WriteBookmarks\relax
%\def\floatpagepagefraction{1}
%\def\textpagefraction{.001}

\shorttitle{}    

\shortauthors{}  

\title [mode = title]{Watt Counts: Energy-Aware Benchmark for Sustainable LLM Inference on Heterogeneous GPU Architectures}  

\tnotemark[1] 

\tnotetext[1]{} 


\author[1]{Mauricio {Fadel Argerich}}%[<options>]
\cormark[1]
\ead{mauricio.fadel@alumnos.upm.es}
\ead[url]{}
\credit{Conceptualization, Methodology, Software, Formal analysis, Investigation, Data curation, Visualization, Writing}

\affiliation[1]{organization={Universidad Politécnica de Madrid},
            addressline={C. de los Ciruelos}, 
            city={Boadilla del Monte},
%          citysep={}, % Uncomment if no comma needed between city and postcode
            postcode={28660}, 
            state={Madrid},
            country={Spain}}

\author[2]{Jonathan Fürst}%[]
% Email id of the second author
\ead{fues@zhaw.ch}
\ead[url]{}
\credit{Methodology, Software, Writing}
\affiliation[2]{organization={Zurich University of Applied Sciences, Switzerland},
            addressline={Gertrudstrasse 15}, 
            city={Winterthur},
%          citysep={}, % Uncomment if no comma needed between city and postcode
            postcode={8400}, 
            state={Zurich},
            country={Switzerland}}

\author[1]{Marta Patiño-Martínez}%[<options>]
\ead{mpatino@fi.upm.es}
\ead[url]{}
\credit{Conceptualization, Supervision, Writing}
\fntext[1]{}


\begin{abstract}
While the large energy consumption of Large Language Models (LLMs) is recognized by the community, system operators lack guidance for energy-efficient LLM inference deployments that leverage energy trade-offs of heterogeneous hardware due to a lack of energy-aware benchmarks and data. In this work we address this gap with Watt Counts: the largest open-access dataset of energy consumption of LLMs, with over 5,000 experiments for 50 LLMs across 10 NVIDIA Graphics Processing Units (GPUs) in batch and server scenarios along with a reproducible, open-source benchmark that enables community submissions to expand this dataset. Leveraging this dataset, we conduct a system-level study of LLM inference across heterogeneous GPU architectures and show that GPU selection is crucial for energy efficiency outcomes and that optimal hardware choices vary significantly across models and deployment scenarios, demonstrating the critical importance of hardware-aware deployment in heterogeneous LLM systems. Guided by our data and insights, we show that practitioners can reduce energy consumption by up to 70\% in server scenarios with negligible impact on user experience, and by up to 20\% in batch scenarios.
\end{abstract}


%\item Dataset with 14M+ power measurements of LLM inference
%\item Benchmark of LLM inference energy across heterogeneous GPUs
%\item Empirical study of 372 LLM–GPU deployments
%\item Analysis of energy consumption in batch and server scenarios

%\nocite{*}

% Keywords

\maketitle

% Main text
\section{Introduction}
While the high-energy demands of Large Language Models (LLMs) are well recognized,  the energy footprint of LLM inference deployments is often overlooked and poorly understood~\cite{jernite2026differentflops}.
Although LLM inference is generally less energy-intensive than training, its total energy consumption is often greater because it requires server deployments with continuous availability to respond to user queries~\cite{you2025how_much_energy_chatgpt}. LLM inference deployments span across a variety of environments including edge servers, on-premise infrastructure, and cloud deployments with heterogeneous hardware, making energy-aware deployment both critical and complex~\cite{yang2024llmcbench}. High energy consumption of these deployments not only generates carbon emissions and increases electricity costs for domestic users and industry~\cite{burian2025_increasing_ai_energy}, but also slows progress towards the Paris Agreement and the UN Agenda 2030 targets on clean energy and climate action~\cite{xue2023strategies}.

In particular, recent studies show that \textit{Graphics Processing Units (GPUs) are the primary energy consumers in AI inference, accounting for up to 90\% of total system power draw}~\cite{argerich2024measuring}, \cite{theodorou2024energy}, underscoring the need for their careful evaluation. This challenge is even greater in modern heterogeneous LLM deployments, where inference workloads are executed on diverse GPU architectures with varying memory capacities, compute characteristics, and power profiles. However, a lack of system-level evidence and tools to benchmark the energy consumption of LLM-GPU configurations means system operators must deal on their own with complex energy-performance trade-offs arising from the added effects of LLM characteristics, GPU architecture, and deployment scenarios.

To address these challenges, we introduce \textit{Watt Counts}, a large-scale, energy-aware benchmark and dataset designed to support deployment decisions for sustainable LLM inference systems. Our contributions are as follows:
\begin{enumerate}
    \item We publish the \textit{largest energy-aware, open dataset of LLMs inference} for system-level analysis of LLM inference across heterogeneous GPU architectures. Our dataset contains power draw, energy per token, throughput, and several additional performance metrics for 50 LLMs running on ten GPUs with  five different NVIDIA architectures in batch and server scenarios. \textit{This results in 5K+ single experiments on 370 different LLM-GPU pairs, and 14M+ power draw samples.} 
    \item A \textit{customizable, open-source, and easy-to-run benchmark for measuring the energy consumption of LLM inference deployments}, enabling reproducible energy and performance measurement in bare metal and cloud instances, which does not require any external power meters. This benchmark allows users to easily contribute to our live dataset. Both the dataset and the source code for the benchmark will be released on Github under an MIT license upon acceptance and are provided in the supplementary material.
    \item A \textit{comprehensive system-level study of LLM inference energy consumption}, revealing how GPU architectural features (e.g., memory bandwidth, cache hierarchy, TDP), model architectural properties, and deployment scenario jointly determine energy efficiency and performance trade-offs. Guided by our data and insights, we show that energy consumption can be reduced by up to 70\% in server scenarios with negligible impact on user experience, and by up to 20\% in batch scenarios.
\end{enumerate}

Together, these resources provide the foundation for data-driven decisions in the deployment of sustainable LLM systems that leverage heterogeneous GPU architectures to improve their scalability and, in particular, their energy efficiency. 

The rest of this paper is organized as follows. Section~\ref{sec:related} surveys related work across related topical areas. Section~\ref{sec:benchmark} describes the benchmark methodology and design, including the energy and performance metrics to be analyzed, and Section~\ref{sec:dataset-construction} describes how the benchmark was used to create the Watt Counts dataset. Section~\ref{sec:analysis} uses the dataset to develop a system-level study of LLM inference energy consumption. Section~\ref{sec:discussion} addresses limitations and broader considerations of this work as well as future work opportunities, and Section~\ref{sec:conclusion} synthesizes the key findings.
\section{Related Work}
\label{sec:related}
The recent widespread use of %large AI models, particularly 
LLMs has made understanding and optimizing their energy consumption a critical research topic. We review previous work in three key areas: (1) energy and carbon measurement studies during training and inference, (2) tools and benchmarks for energy estimation, and (3) energy-aware datasets for LLM inference. 

% Analyzing energy consumption of LLMs in training
\textbf{Energy and carbon measurement studies during training and inference.} Several works have focused on measuring the energy consumption and carbon emissions of LLMs during training~\cite{liu2024green, patterson2021carbon, mcdonald2022great}, as this phase is typically the most energy-intensive.
% Analyzing energy consumption of LLMs in inference
However, the inference phase often has a  greater impact than training, as LLMs spend most of their operational lifetime serving inference requests~\cite{samsi2023words, bai2024beyond, morrison2025holistically, argerich2024measuring}. In this line, Wu et al. analyze the environmental impact of experimentation, training, and inference of LLMs and recommendation models at Meta~\cite{wu2022sustainable}. Jegham et al. estimate the energy consumption of several proprietary LLMs based on company-disclosed information since no energy-aware data is available for these models~\cite{jegham2025hungry}. Similarly, Kim et al. survey numerous optimization techniques for transformers, evaluating their impact on inference time and energy, and applying them to improve energy efficiency~\cite{kim2023full}. MLPerf Power~\cite{tschand2025mlperf} presents insights from over 1800 energy consumption measurements of various ML workloads, including LLMs, across different systems. These were obtained by running the MLPerf Inference benchmark~\cite{reddi2019mlperf} and recording system-level energy consumption with physical power meters. TokenPowerBench~\cite{niu2025tokenpowerbench}, a benchmark for measuring phase-level power consumption of LLM inference, focuses on model and inference engine comparison, evaluating 15+ open-source models only on H100 clusters for batch scenarios. At the time of writing, the dataset and benchmark implementation of TokenPowerBench are not public.

% Tools/benchmarks to measure energy consumption.
\textbf{Tools and benchmarks for energy estimation.} In many settings, such as cloud computing, inference runs on systems that cannot be instrumented with physical power meters. To address this, several software tools have been developed, including Carbontracker~\cite{anthony2020carbontracker} and CodeCarbon~\cite{schmidt2021codecarbon}, which estimate carbon emissions, and EnergyMeter~\cite{argerich2024measuring}, which estimates energy consumption of virtualized environments or bare metal instances. Zeus~\cite{zeus-nsdi23} offers a library for measuring and optimizing energy consumption of ML training workloads by adjusting parameters such as power limits and batch size. Recently, ML.ENERGY~\cite{mlenergy-neuripsdb25} presents an LLM energy efficiency leaderboard to collect energy per response for different LLMs and tasks, based on measurements collected with Zeus. In addition, recent works have explored the performance and energy characteristics of LLM workloads. Frameworks such as DynamoLLM~\cite{stojkovic2025dynamollm}, LLMCO2~\cite{fu2025llmco2}, throttll'em~\cite{kakolyris2025throttll}, and Patel et al.~\cite{patel2024characterizing} analyze and aim to optimize energy or carbon footprint for specific GPUs (A100, H100) and models. Fernandez et al. introduce a modeling approach to estimate the energy efficiency of LLM inference, testing it on three GPUs and seven LLMs using PyTorch and vLLM~\cite{fernandez2025energy}. Similarly, workload-based analyses on multi-GPU systems~\cite{wilkins2024offline,kakolyris2025throttll,agrawal2024vidur} focus on optimization for specific hardware. Our dataset provides empirical data across a much wider hardware space to complement and enable the evaluation of these modeling approaches.

% Energy-aware LLM datasets.
\textbf{Energy-aware datasets for LLM inference.} Table \ref{tab:dataset_comparison} highlights a key gap in the literature: there is currently no open dataset or benchmark that enables systematic characterization of LLM inference energy across a diverse set of models, heterogeneous GPUs, and batch and server scenarios.
Most existing studies are limited in scope, covering only a unique or narrow range of GPUs~\cite{husom2024price, guo2025estimating, mlenergy-neuripsdb25, aienergyscore-leaderboard}, are restricted to batch scenarios, or in the case of MLPerf Power~\cite{tschand2025mlperf}, they do not report GPU-level energy measurements, instead providing only aggregate system-level metrics. 

\begin{table*}[ht]
  \caption{Comparison of energy-aware LLM inference datasets 
  %and studies
  }
  \label{tab:dataset_comparison}
  \centering
  \resizebox{1.0\hsize}{!}{
  \begin{tabular}{p{6cm}ccccp{2.5cm}cp{3cm}}
    \toprule
    \textbf{Dataset} &
    \textbf{\# GPUs} &
    \textbf{\# LLMs} &
    \textbf{LLM-GPU comb.} &
    \textbf{Energy Metrics} &
    \textbf{Workload Type/Scenario} &
    \textbf{Open Data} &
    \textbf{Engine} \\
    \midrule
    LLM Energy Consumption Dataset~\cite{husom2024price} &
    3 &
    6 &
    18 &
    \textcolor{teal}{\cmark} &
    Batch & 
    \textcolor{teal}{\cmark}  &
    ollama \\
    \midrule
    Azure LLM Inf. Traces~\cite{stojkovic2025dynamollm} &
    1 &
    6 &
    6 &
    \textcolor{red}{\xmark$^1$} &
    Server & 
    \textcolor{teal}{\cmark} &
    vLLM \\
    \midrule
    %LLM Energy Cons. Dataset~\cite{nitishkumar2k01_llms_energy_dataset_2025} & 4 & 27 & \textcolor{red}{\xmark} & \textcolor{teal}{\cmark} & Training & \textcolor{teal}{\cmark} & Unknown \\ \midrule
    AI Energy Score~\cite{aienergyscore-leaderboard} &
    1 &
    121 % 184 models in total (image, ASR)
    &
    121 &
    \textcolor{teal}{\cmark} &
    Batch & 
    \textcolor{teal}{\cmark} &
    PyTorch, vLLM, Optimum$^2$
    \\
    \midrule
    ML.ENERGY~\cite{mlenergy-neuripsdb25} &
    2 &
    27 % more models for video, image
    &
    54 &
    \textcolor{teal}{\cmark} &
    Batch & 
    \textcolor{teal}{\cmark} &
    vLLM
    \\
    \midrule
    MLPerf Power~\cite{tschand2025mlperf} &
    11$^3$ &
    3 &
    33 &
    \textcolor{orange}{\cmark$^4$} &
    Server, Batch, Single/Multiple Stream & 
    \textcolor{teal}{\cmark} &
    PyTorch, ONNX
    \\
    \midrule
    J. Fernandez et al.~\cite{fernandez2025energy} &
    3 &
    7 &
    21 &
    \textcolor{teal}{\cmark} &
    Batch & 
    \textcolor{red}{\xmark} &
    PyTorch, vLLM
    \\
    \midrule
    \textbf{Watt Counts (ours)} &
    \textbf{10} &
    \textbf{50} &
    \textbf{372} &
    \textcolor{teal}{\cmark} &
    Server, Batch & 
    \textcolor{teal}{\cmark} &
    vLLM
    \\
    \bottomrule
  \end{tabular}}
  \scriptsize
  \raggedright
  $^1$Although the dataset does not publish per-run energy measurements, the paper reports aggregated energy statistics. \\
$^2$Although these inference engines are listed as supported, the published data does not specify which inference engine was used; it is therefore assumed that the default PyTorch engine was employed. \\
$^3$We consider only GPUs used for LLM workloads; devices such as the Cloud AI 100 Pro and Jetson AGX Orin are excluded. Including these devices would increase the total number of evaluated devices by MLPerf to 24. \\
$^4$MLPerf Power reports energy consumption at the system level only; GPU-level energy measurements are not provided.
\end{table*}

% Why we are different
No publicly available dataset spans a diverse set of both models and heterogeneous GPU architectures under both batch and server scenarios. This gap is consequential for the deployment of heterogeneous system for LLM inference. Without cross-LLM-GPU empirical data, system operators cannot make informed decisions about GPU selection, workload placement, or energy-performance trade-offs in heterogeneous LLM deployments. \textit{Watt Counts} addresses this directly: it is the first large-scale, open dataset and benchmark covering 50 LLMs across 10 GPUs spanning five NVIDIA architectures, in both batch and server scenarios, fully reproducible without physical power meters. It provides the empirical foundation that the field lacks for data-driven, hardware-aware deployment of LLMs at scale.

%Prior studies have examined LLM energy consumption in limited or isolated settings, and existing benchmarking tools often require physical access or time-consuming profiling. Our work builds upon and complements these efforts by providing the first large-scale, open dataset and benchmark for LLM inference energy that spans diverse models, architectures, and GPUs. Unlike prior studies, our benchmark is fully reproducible on both bare-metal and cloud instances without requiring physical power meters, and the dataset captures both batch and server scenarios. This enables systematic characterization of LLM energy consumption, facilitates comparison across hardware and models, and provides a foundation for future research on energy-efficient deployment of LLMs.

\section{Benchmarking Methodology}
\label{sec:benchmark}
This section describes the design of our benchmark and its use for the construction of the Watt Counts dataset. 
We first present the research questions that motivate the benchmark design, dataset construction and analysis (Section~\ref{subsec:research-questions}). 
Next, we introduce the design of the benchmark, including the defined inference scenarios, the energy and performance metrics collected, and its implementation to ensure the experiments are executed in a reproducible manner (Section~\ref{subsec:benchmark-design}). 

\subsection{Research Questions}
\label{subsec:research-questions}
To systematically examine how energy efficiency varies with respect to GPU characteristics, LLM properties, and workload scenarios, we define the following research questions:

\begin{enumerate}[label=\textbf{RQ\arabic*:}, leftmargin=*, itemsep=4pt]
    \item \textbf{GPU Architecture Efficiency Across LLMs}. Is a single GPU consistently the most energy-efficient across all LLMs, or do efficiency rankings vary? Which GPU architectural characteristics (e.g., memory capacity, memory bandwidth, FLOPS, TDP) most strongly correlate with energy efficiency for LLM inference?
    \item \textbf{LLM Efficiency Across GPUs}. 
    What model properties (such as size, model family, etc.) impact energy efficiency significantly during inference across GPUs?
    \item \textbf{Deployment Scenario Sensitivity}. How does the deployment scenario (batch vs. server load) influence the energy efficiency of LLMs on a given GPU? Are the same LLM–GPU pairs optimal across server and batch?
\end{enumerate}

These questions guide the benchmark design, its use to construct the Watt Counts dataset, and the subsequent analysis of LLM energy efficiency.

\subsection{Benchmark Design}
\label{subsec:benchmark-design}
The benchmark defines LLM inference for Question-Answering (QA) as the application under test, a common use case for LLMs. This application is used in two predefined scenarios (batch and server), and during execution, a metrics collector captures energy and performance measurements.
The benchmark design is guided by three principles: reproducibility, modularity, and practical relevance. To support community adoption and extension of our benchmark and dataset, both are open-source and publicly available.

\begin{figure}[t]
    \centering
\includegraphics[width=1.0\linewidth]{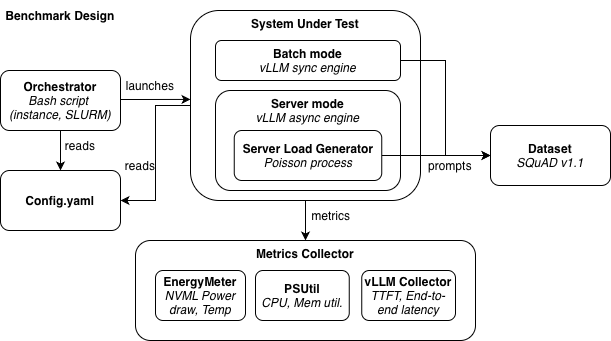}
\caption{Benchmark design overview. The orchestrator launches the System Under Test (SUT) based on a user-provided configuration. The SUT runs in batch mode (vLLM synchronous engine) or server mode (vLLM asynchronous engine with a Poisson-based load generator). The metrics collector aggregates energy, system, and LLM-level performance measurements throughout execution.}
\label{fig:benchmark-design}
\rmspace
\end{figure}

The benchmark is implemented in Python and structured in four main components as shown on Figure~\ref{fig:benchmark-design}. The \textit{orchestrator} reads a user-provided \texttt{config.yaml} specifying the models (via their Hugging Face hub handle), deployment scenarios, number of iterations, and request arrival rates, and launches the System Under Test (SUT.) The SUT is a vLLM inference engine~\cite{kwon2023efficient} operating in one of two modes: \textit{batch}, which uses the synchronous engine to process all prompts in a single submission, and \textit{server}, which uses the asynchronous engine paired with a \textit{server load generator} that dispatches prompts following a Poisson arrival process. Throughout execution, the \textit{metrics collector} records GPU metrics using EnergyMeter, CPU and memory metrics with PSUtil, and collects LLM-level performance indicators including TTFT and end-to-end latency from the vLLM metrics. Additionally, before each run, detailed information about the underlying hardware, including CPU, GPU, memory, and disk specifications are collected. This metadata ensures reproducibility and allows the analysis of how hardware characteristics influence energy and performance metrics. 

vLLM was selected as it is often at the top of latency, throughput, and energy performance studies when compared to other inference engines such as SGLang or TensorRT-LLM~\cite{fernandez2025energy, stuhlmann2025bench360} and is considered one of the most  complete and widely used general purpose inference engines~\cite{park2025survey}. Its main optimization is PagedAttention, which enables a fine-grained GPU memory allocation and reuse to maximize throughput and minimize memory usage. 
To support setups from single-node deployments to large-scale evaluations on HPC clusters, we provide bash scripts for orchestrating benchmark runs on bare-metal and cloud instances, as well as SLURM-managed clusters.

To improve measurement accuracy and reproducibility, we control sources of non-determinism by fixing relevant parameters (i.e., temperature$=0$, top\_k$=0$, top\_p$=1$, maximum generation length$=256$, repetition\_penalty$=1$) and random seeds (SamplingParams, Python's random, numpy's random, all Pytorch's seeds), acknowledging that full determinism is not achievable in practice unless using specialized libraries, due to GPU floating-point non-determinism and dynamic kernel scheduling~\cite{gond2026llm}. We utilize default settings for vLLM including its GPU memory utilization of 90\%, except for GPUs with less than 20GB where we set it to 80\% to avoid out of memory errors that occur otherwise. The maximum context length is set to 1024 and the maximum generation length to 256 tokens to reflect typical QA interaction lengths while keeping total benchmarking time tractable across 370+ LLM-GPU configurations.

\subsubsection{Deployment Scenarios}
\label{subsec:scenarios}
The benchmark evaluates LLM inference for two representative deployment scenarios, based on 
MLPerf Inference~\cite{reddi2020mlperf}: 

\begin{enumerate}

\item \emph{Batch Processing (Offline Processing).} This scenario represents the use of LLMs in high-throughput scientific and data analysis workflows, where a large number of queries are processed in bulk. For example, LLMs are extensively applied to Information Extraction (IE), where documents are processed in a batch to extract relevant structured information~\cite{colakoglu-etal-2025-problem}. In this scenario, 1,000 queries are submitted simultaneously to the vLLM engine in synchronous mode. Prompts are drawn from the SQuAD v1.1 dataset~\cite{rajpurkar2016squad}, which contains crowd-sourced questions based on Wikipedia articles and is commonly used to evaluate QA systems. Energy is measured and attributed only to the execution of the finite workload, i.e., idle periods before or after batch execution are outside the scope of this scenario, as it models a bounded computational task rather than a continuously running service.

\item \emph{Server Mode (Online Processing).} This scenario simulates real-world LLM applications, where the model is deployed as a service and must handle several requests concurrently. The load in the server is variable as it depends on the number of concurrent users and queries the application receives. 
In this scenario, requests using prompts from the SQuAD v1.1 dataset are sent to the asynchronous vLLM engine, which handles concurrent requests without blocking, processing requests as they arrive following a Poisson arrival process (as in MLPerf~\cite{reddi2019mlperf}), with rates $\lambda=\{0.017, 0.3\}$ corresponding to low-load and high-load conditions, respectively. 

%\begin{equation}
%P(k; \lambda) = \frac{\lambda^k e^{-\lambda}}{k!}
%\end{equation}
%$P(k; \lambda)$ represents the probability of observing exactly $k$ user requests in a given interval of time, for a process with an average of $\lambda$ requests per interval. $k$ is the number of %occurrences of requests, and $e$ is Euler's number. 

\end{enumerate}

Each scenario includes an initial warmup phase: for the batch scenario, this is equal to running inference for one batch, for server, the warmup runs the scenario for one minute. This warmup is followed by five independent iterations per configuration (batch, low-load server, and high-load server).
After the warmup and between iterations, we introduce a cold-down period in which the system remains idle to stabilize hardware conditions: experiments start once the GPU power draw remains within a 3 W range for at least 30 seconds -- empirically determined to reliably indicate that residual heat from the previous iteration no longer influences power measurements -- and the GPU temperature falls below 65 °C, ensuring that thermal throttling, which normally occurs from 75 °C, does not influence the results. In addition, we enforce a 5 minute timeout per cold-down period to avoid blocking the benchmark in cases in which the temperature does not drop from its defined threshold. In the benchmark, each server iteration runs for five minutes, a duration empirically shown to yield energy measurements within <2\% deviation (95\% confidence interval) when compared to one-hour runs, while substantially reducing total benchmarking time. 

%Both scenarios exhibit different energy consumption patterns and are influenced by different factors. In the Server scenario, the instance must remain available for extended periods, even when no requests are incoming, making the GPU idle power consumption and its behavior to manage bursty loads decisive factors in the overall energy consumption. In contrast, Batch processing is throughput-bound: the workload has a finite, defined scope, and total energy consumption is directly tied to how quickly the task can be completed. 

\subsubsection{Energy Metering} 
\label{subsubsec:energy-metrics}
We leverage EnergyMeter~\cite{argerich2024measuring} to measure the energy consumption of the GPU through pyNVML, which obtains the GPU power draw using the NVIDIA Management Library (NVML) and integrates it over execution time to obtain energy consumption in Joules. NVML was chosen because it is natively supported by NVIDIA GPUs in both bare metal and virtualized environments, enabling consistent and reproducible measurements across heterogeneous NVIDIA GPUs and deployment settings without external instrumentation. This significantly increases the portability of the benchmark, enabling the addition of more data to our dataset in the future. While NVML does not provide circuit-level power measurements, it allows to separate the GPU energy from other components. Moreover, prior work has shown its measurements deviations from external power meters are below $5\% $~\cite{yang2024accurate}, which is sufficient for comparing performance across models and GPUs, the main goal of our benchmark. 

GPU power draw is sampled with NVML at board level at 10 Hz during each experiment, reflecting the complete device under exclusive use; each measured device corresponds to a single physical GPU with its own framebuffer memory (in the case of NVL devices, 94~GB for the H100 NVL and 141~GB for the H200 NVL, \texttt{MultiGPU Board: No} in both cases). Sampling frequencies ranging from 1 Hz to 200 Hz were evaluated empirically and 10 Hz was selected as it was observed to provide the best trade-off between measurement overhead and accuracy, with an integrated energy error of <1\% when compared to 200 Hz sampling and an execution time overhead of <0.5\%. 

We focus on GPU energy consumption, as it has been shown to be the dominant energy consumer in LLM inference. To validate this assumption, we ran a subset of our experiments on a representative bare metal instance (Intel(R) Core(TM) i9-9900K @ 3.60GHz, 32 GB RAM, a single NVIDIA GeForce RTX 4090) using EnergyMeter to collect energy metrics for the GPU via NVML, and the CPU and DRAM via RAPL. In batch and high load server scenarios, the GPU consumed 91.2-92.5\% of the total energy consumed by the three components, while in the low load server scenario it represented ~78.7\% and the CPU ~16.6\%, due to the GPU being idle for large periods of time. In all scenarios, the GPU remained the main energy consumer, exhibiting a 0.998 correlation with the total measured energy consumption. 

\subsubsection{Energy Metrics}
The total energy consumed by the GPU during LLM inference is given by: $E=P \cdot t$ 
where $P$ denotes the average power draw during inference, and $t$ is the measurement period. However, total energy consumption is often an impractical measurement for comparisons across LLM deployments, as it depends on specific workload factors including input size and execution duration, which motivates the use of normalized energy efficiency metrics.

%\paragraph{Batch scenarios}
In \textbf{Batch scenarios}, the total energy consumption is determined by the interplay of power draw and execution time. While higher throughput, i.e., the number of tokens generated per unit of time, reduces execution time, lower power draw can also reduce total energy even at reduced throughput~\cite{maliakel2025characterizing}, making \textit{energy per token} the appropriate metric as it jointly captures both dimensions. 

In \textbf{Server scenarios}, the system must remain available to handle incoming requests, so total execution time is largely decoupled from the system's processing speed. Moreover, the system might spend long periods idling, waiting for incoming requests. As a result, attributing total energy consumption to generated tokens or individual requests requires assumptions about request arrival rates and idle power consumption, making energy per token highly deployment-dependent in this context. We therefore utilize \textit{mean power draw}, as it captures idle and active periods and enables consistent comparisons of LLM-GPU deployments in server scenarios.

\subsubsection{Additional Performance Metrics}
In addition to energy metrics, we collect a set of system-level and LLM-level performance metrics at a 10 Hz frequency. System metrics include GPU utilization, GPU memory usage, GPU temperature, GPU clock frequency, CPU utilization, and main memory usage. LLM-level metrics include number of input tokens (per request and in total), number of tokens generated (per request and in total), the total duration of the experiment, Time-To-First-Token (TTFT), queuing time and end-to-end response time. These measurements enable the computation of additional metrics and statistics, including throughput and aggregated performance indicators such as response-time percentiles, useful for balancing energy consumption with service-level constraints, such as response time or throughput requirements.

Since the benchmark is designed to characterize energy and performance independently of task-specific outputs, accuracy evaluation is intentionally excluded from scope; the systems output correctness does not affect power draw, energy consumption, latency nor throughput measurements. Practitioners are referred to the respective model documentation for task-specific accuracy benchmarks.

\section{Dataset Construction}
\label{sec:dataset-construction}
We leverage our benchmark to create a comprehensive dataset that captures energy consumption patterns %and performance metrics 
for a wide range of LLM-GPU pairings. 

We select 50 open-source, popular models from the Hugging Face Hub, covering a wide range of architectures, sizes, and vendors. Because our objective is to characterize LLM energy and performance across GPUs, we limit model size to 30B parameters (requiring approximately 60GB in 16-bit precision) to ensure that each model fits on at least two of the evaluated GPUs. The current version of the dataset focuses on single-GPU setups to thoroughly evaluate model power behavior across diverse GPU architectures, avoiding additional noise and variability introduced by factors such as communication overhead, tensor-parallel and pipeline-parallel synchronization costs, memory sharding strategies, and interconnect bandwidth limitations.

For hardware, we select a wide range of NVIDIA GPUs, spanning five different architectures and including server-class and consumer-grade devices (Table~\ref{tab:gpus}), enabling the analysis of architectural design trade-offs that affect energy efficiency. The selected GPUs vary significantly in memory capacity (16GB–141GB) and thermal design power (TDP) (70W–600W), reflecting the heterogeneity of real-world LLM deployment environments.

% 29.12.2025: MFA verified all specs, added Profile and L2 Cache, and added more precision to GPU names. Minor changes in values were also updated so it fits values reported by NVIDIA.
\begin{table*}[ht]
    \centering
    \caption{NVIDIA GPUs used in the dataset and their main technical characteristics: architecture, profile (consumer or enterprise), memory, TDPs, tensor FP16 TFLOPS without sparsity, memory bandwidth, L2 cache size, TFLOPS (without sparsity) per Watt, and release year.}
    \label{tab:gpus}
    \resizebox{1.0\linewidth}{!}{%
    \begin{tabular}{lllrrrrrrr}
    \toprule
    GPU & Architecture & Profile & \shortstack{Memory\\(GB)} & TDP (W) & \shortstack{TFLOPS\\(FP16)} & \shortstack{Mem BW\\(GB/s)} & \shortstack{L2 Cache\\(MB)}& \shortstack{TFLOPS\\per Watt} & Year \\
    \midrule
    Tesla V100 SXM2 & Volta & Enterprise & 32 & 250 & 125 & 898 & 6 & 0.5 & 2018 \\
    Tesla T4 & Turing & Enterprise & 16 & 70 & 65 & 320 & 4 & 0.93 & 2018 \\
    A100 SXM4 & Ampere & Enterprise & 40 & 400 & 312 & 1555 & 40 & 0.78 & 2020 \\
    GeForce RTX 3090 & Ampere & Consumer & 24 & 350 & 71 & 936 & 6 & 0.20 & 2020 \\
    A30 PCIe & Ampere & Enterprise & 24 & 165 & 165 & 933 & 24 & 1.00 & 2021 \\
    GeForce RTX 4090 & Ada Lovelace & Consumer & 24 & 450 & 165 & 1010 & 72 & 0.37 & 2022 \\
    L40S & Ada Lovelace & Enterprise & 48 & 300 & 362 & 864 & 48 & 1.21 & 2022 \\
    L4 & Ada Lovelace & Enterprise & 24 & 72 & 121 & 300 & 48 & 1.68 & 2023 \\
    H100 NVL & Hopper & Enterprise & 94 & 400 & 835 & 3940 & 50 & 2.39 & 2023 \\
    H200 NVL & Hopper & Enterprise & 141 & 700 & 835 & 4800 & 50 & 1.19 & 2024 \\
    \bottomrule
    \end{tabular}
    }
\end{table*} 

%Each of these combinations constitutes a single experimental configuration, which we repeat five times. Overall, this results in 5000 successful experimental configurations for 137/150 models out of overall 22500 theoretical possible runs (success rate: 59\%). Most failures stemmed from GPU memory limitations or framework incompatibilities. \todo{Add information about the issues.}
%The majority of failed model runs is due to insufficient GPU memory on smaller GPUs (e.g., T4). However, we also encountered other incompatibility errors for specific vLLM-GPU-LLM combinations. %(see supplementary material for details).

\section{Dataset Analysis and Insights}
\label{sec:analysis}
This section presents a systematic analysis based on the Watt Counts dataset described in Section~\ref{sec:dataset-construction}, comprising 5,000+ experiments across 50 LLMs and 10 GPUs in batch, low-load server, and high-load server scenarios. Each configuration was repeated five independent times following the benchmark methodology described in Section~\ref{sec:benchmark}.

\subsection{RQ1: GPU Architecture Efficiency Across LLMs}
\label{subsec:rq1}
\textit{Is a single GPU consistently the most energy-efficient across all LLMs, or do efficiency rankings vary? Which GPU architectural characteristics (e.g., memory capacity, memory bandwidth, FLOPS, TDP) most strongly correlate with energy efficiency for LLM inference?}

We measure the energy per token each model achieves across GPUs in the batch scenario, ranking the GPUs from the most efficient, i.e., the one that achieves the lowest energy per token for the model, to the least efficient. 
In particular, we analyze batch inference data because this scenario minimizes idle periods (idle periods average 2\% in the dataset), providing a cleaner signal for correlating GPU architectural characteristics with energy efficiency. The impact of workload scenario (batch vs. server) is examined separately in RQ3.
We summarize the rankings for all models in Table~\ref{tab:gpu_ranks_combined}, together with the total number of models executed on each GPU -- determined by  memory capacity constraints and compatibility limitation on older devices (Tesla V100, T4.)

Across all models, the H100 achieves the lowest energy per token in ~90\% of cases.
The H200 ranks second for most models and only achieves first place for two of the largest models in our set (`google/gemma-3-27b-it`, `nvidia/NVIDIA-Nemotron-3-Nano-30B-A3B-BF16`.) In contrast, the L4 achieves the first place for three models (`Qwen/Qwen2.5-0.5B-Instruct`, `openai-community/gpt2`, `tiiuae/Falcon3-Mamba-7B-Instruct`,) two of which are among the smallest. This suggests that energy efficiency depends on interactions between GPU characteristics and model size.

\begin{table}[t]
\centering
\caption{GPU ranking positions of energy-per-token achieved in the batch scenario. Only GPUs ranked 3rd or above for at least one model are shown.}
\label{tab:gpu_ranks_combined}
\resizebox{1.0\linewidth}{!}{%
\begin{tabular}{lrrrr}
\toprule
 GPU                   &   1st &   2nd &   3rd &   Total models \\
\midrule
A100 SXM4 40 GB  &     0 &     0 &    13 &             42 \\
 A30 PCIe         &     0 &     1 &    13 &             34 \\
 GeForce RTX 4090 &     0 &     1 &     1 &             34 \\
 H100 NVL 94 GB   &    45 &     4 &     1 &             50 \\
 H200 NVL         &     2 &    43 &     2 &             50 \\
 L4               &     3 &     1 &    11 &             33 \\
 L40S             &     0 &     0 &     2 &             43 \\
\bottomrule
\end{tabular}
}
\end{table}

To analyze this interaction, we group the models in the dataset into four categories based on their hardware compatibility observed during benchmarking (FP16 inference, fixed context length). Specifically, we define the following size categories:
\begin{itemize}
    \item Small: models that fit on all evaluated GPUs, comprising models between 0.124B and 6B parameters.
    \item Medium: models that fit on all GPUs with more than 16GB of memory (i.e., excluding the T4 in our setup). This includes models with 6B-10.7B parameters.
    \item Large: models that fit on GPUs with more than 32GB (i.e., excluding the T4, A30, RTX 3090, RTX 4090, and L4), including models with 12B-20B parameters.
    \item Xlarge: models that fit on GPUs with more than 48GB of memory (i.e., H100 and H200), in our setup this includes models with 20B-27B parameters.
\end{itemize}

\begin{figure}[ht]
    \centering
\includegraphics[width=1.0\linewidth]{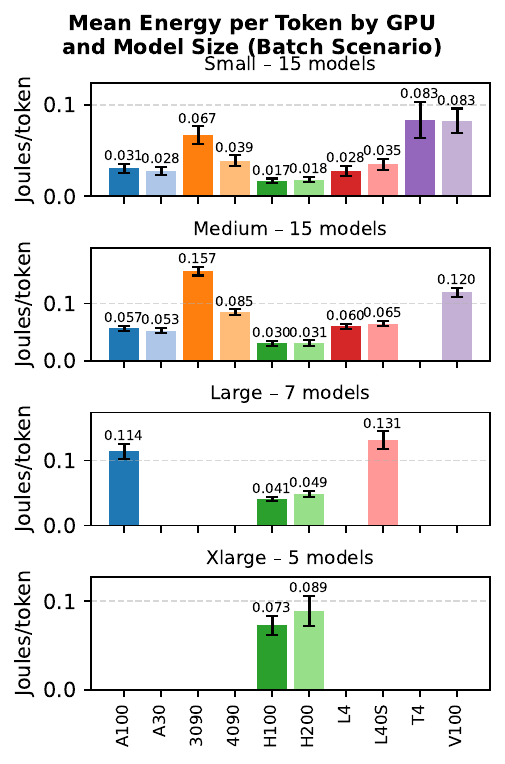}
\caption{Mean energy per token for model size categories across GPUs for selected models. Error bars show 95\% confidence intervals. Missing bars indicate that the GPU could not run any models in that size category due to memory or compatibility limitations.}
\label{fig:energy_per_token_sizes}
\rmspace
\end{figure}

Models are assigned to the smallest compatible category based on whether they successfully loaded without out-of-memory errors. We compute the mean energy per generated token for each GPU and category as shown in Figure~\ref{fig:energy_per_token_sizes}. To ensure fair comparisons, each category includes only dense models that ran on all GPUs in that category.

The H100 achieves the lowest mean energy per token across all categories, followed by the H200.
Interestingly, for small and medium models the A30 achieves third place, tied with the L4 for small models, both surpassing more powerful GPUs in terms of TFLOPS and memory bandwidth. %This result is noteworthy given that the L4 offers the second lowest tensor computing performance (121 TFLOPS) and the lowest memory bandwidth (300 GB/s), and the A30 also has a rather mediocre performance (165 TFLOPS and 933 GB/s), which translates them into the second lowest throughput for the L4 and a mid-rank throughput for the A30, as shown in Figure~\ref{fig:throughput_sizes}. 
Their advantage is explained by their low TDPs, which compensate for their low throughput. %The A30 has a higher TDP (165W,) but this is still low when compared to the other GPU's TDPs that reach up to 600W. However, TDP cannot be used to estimate the energy efficiency of a GPU and the T4 exemplifies this: even though it has the lowest TDP at 70W, it is the least efficient GPU for the small category, affected by achieving the lowest throughput.

\begin{figure}[ht]
    \centering
\includegraphics[width=1.0\linewidth]{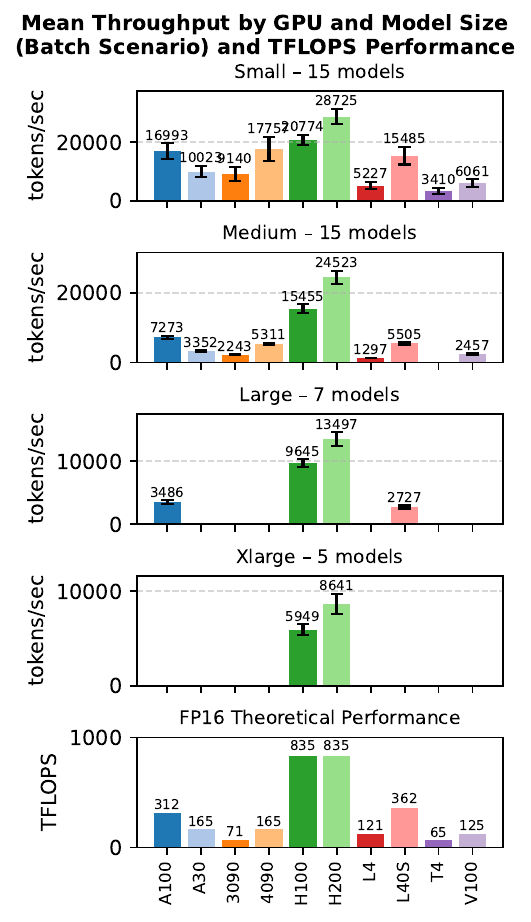}
\caption{Mean throughput for model size categories across GPUs for selected models. Error bars show 95\% confidence intervals. Missing bars indicate that the GPU could not run any models in that size category due to memory or compatibility limitations.}
\label{fig:throughput_sizes}
\rmspace
\end{figure}

For throughput, the H200 achieves the highest mean in all categories (Figure~\ref{fig:throughput_sizes}), with the H100 achieving a significantly lower throughput (38\%-58\% lower than the H200's.) This advantage likely stems from the H200's higher memory bandwidth, a 25\% higher than the one of H100. A similar behavior is observed in the L40S and the A100, while the L40S has a ~16\% higher TFLOPS performance, the A100 achieves a higher throughput in all categories thanks to its ~80\% higher memory bandwidth. This is consistent with previous work that has identified memory bandwidth as a primary bottleneck in LLM inference~\cite{recasens2025mind}.

These results show that TFLOPS performance does not imply higher throughput, and a higher throughput does not necessarily translate to lower energy per token. For instance, deploying an Xlarge model in an H100 can save 20\% of energy compared to the H200, which would be selected if we were to optimize the throughput. Moreover, a commonly used energy efficiency proxy, TFLOPS per Watt, is a poor predictor at the system level across full inference workloads: as shown on Table~\ref{tab:gpus} and Figure~\ref{fig:energy_per_token_sizes}, GPUs with higher TFLOPS/W do not consistently achieve lower energy per token.

\textbf{Answering RQ1:} \textit{While no single GPU is universally the most energy-efficient across all LLMs, the H100 dominates in 90\% of models in the batch scenario.} Exceptions arise for smaller and very large models, suggesting that GPU efficiency rankings are largely consistent but sensitive to model size. 
GPUs with lower TDP and moderate performance, such as the L4 and the A30, can achieve a lower energy per token than more powerful GPUs for running inference with small models, while large models take advantage of powerful GPUs, offsetting their higher power draw with high throughput. %In addition, across a wide range of models, a balanced GPU, such as the H100, with high memory bandwidth, substantial compute capacity and a moderate power draw, achieves the best energy efficiency. 
Overall, energy efficiency depends on the specific LLM-GPU combination, with model properties interacting with GPU architecture properties; mainly TDP, memory bandwidth, and TFLOPS. These findings demonstrate that energy-efficient LLM inference depends on hardware-aware model placement, motivating further analysis of how model architectural properties affect energy efficiency.

\subsection{RQ2: LLM Efficiency Across GPUs}
\label{subsec:rq2}
\textit{What model properties (such as size, model family, etc.) impact energy efficiency significantly during inference across GPUs?}

RQ2 examines how LLM architectural properties affect energy efficiency across GPUs. As in RQ1, we analyze batch inference data in RQ2, as it provides a cleaner signal for correlating model architectural properties with energy efficiency. RQ1 has already identified model size as a factor that affects energy efficiency and throughput across GPUs (Figures \ref{fig:energy_per_token_sizes}, \ref{fig:throughput_sizes}). When looking at each LLM-GPU combination, energy per token strongly varies spanning nearly three orders of magnitude (from $0.003J$ to $1J$). 

\begin{figure}[t]
    \centering
\includegraphics[width=1.0\linewidth]{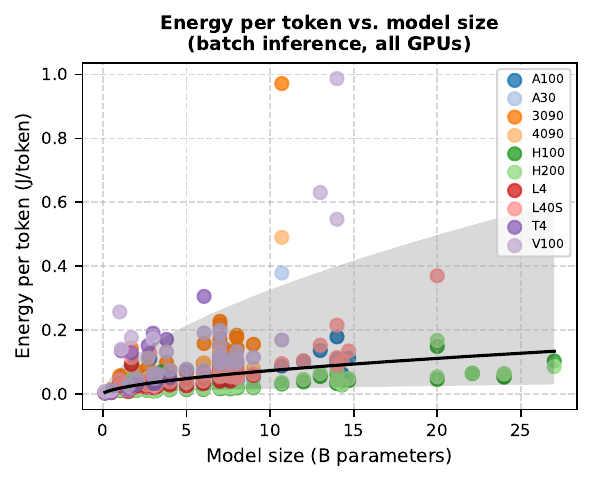}
\caption{The gray area shows the 95\% prediction interval of a log–log regression, $\log(E)=\alpha\log(\text{size})+\beta$, fitted over all data points; data points outside this envelope are considered energy outliers.}
\label{fig:energy_per_token_vs_size}
\rmspace
\end{figure}

Figure~\ref{fig:energy_per_token_vs_size} shows the mean energy per token against the number of \emph{active parameters} for each LLM--GPU pair across all evaluated configurations. We use active rather than total parameters to ensure comparability between dense and mixture-of-experts (MoE) architectures, as energy consumption during inference is driven by the parameters involved in the operations per token rather than the full model capacity. For dense models, active and total parameters are equivalent. For MoE models, active parameters correspond to the number of parameters engaged per forward pass -- i.e., the parameters of the selected experts plus the shared non-MoE layers -- as reported in the respective model cards or original publications. The full list of models with their total and active parameter counts is provided in Table~\ref{tab:model_list}.

Previous works~\cite{argerich2024measuring, niu2025tokenpowerbench} have observed a sublinear relationship in the increase of energy with respect to active parameters, so we fit a log--log regression line to capture this relationship in Figure~\ref{fig:energy_per_token_vs_size}, with the grey area showing the 95\% prediction interval of the fitted log--log regression. The remaining outliers arise from implementation-specific (e.g., overflow behavior or reliance on ad hoc library versions) or hardware-related factors (limited support in the Tesla V100 and T4 for modern fused kernels leads to higher energy consumption.) More details are provided in the appendix.

We further analyze the effect of model size and architectural features on energy with two mixed-effects models: Model A quantifies the effect of model size on energy, and Model B adds architectural features to isolate their independent contributions beyond size. 

\paragraph{Model A.} To quantify the effect of size on energy consumption, we use a mixed-effects model with a random intercept per GPU, regressing $\log(E_{i,j})$ on $\log(P_i)$ while controlling for model type (i.e., model family architecture.) The results reveal a strong and highly significant scaling relationship, where the coefficient for $\log(P_i)$ is $0.237$ ($p < 0.001$), i.e., a 10$\times$ increase in active parameters approximately increases $1.7\times$ the energy per token.

\paragraph{Model B.} Furthermore, Figure~\ref{fig:energy_per_token_vs_size} shows that many models of similar size exhibit substantially different energy per token even on the same GPU, suggesting that architectural properties beyond parameter count influence energy consumption. To quantify their effects while accounting for systematic differences across GPUs, we use another mixed-effects model. Based on prior work~\cite{argerich2024measuring, kim2023full}, we focus on a set of architectural properties expected to have a first-order impact on energy usage: (i) \textit{key-value head count}, i.e., the number of heads in the key-value cache of the attention mechanism, (ii) \textit{attention head count}, i.e., the number of parallel attention heads, (iii) \textit{hidden size}, i.e., the width of the model's internal representations, and (iv) \textit{number of layers}, i.e., the model depth.

However, architectural features such as depth, width, and attention structure are strongly correlated with model size, so directly including them alongside $\log(P_i)$ would introduce multicollinearity. To isolate the effect of architectural properties independently of size, we residualize each numerical property with respect to $\log(P_i)$ before running the mixed-effects model. Concretely, for each architectural feature $Z_i$, we estimate a linear regression of the form
\[
Z_i = \alpha_0 + \alpha_1 \log(P_i) + r_i,
\]
and use the residual $r_i$ -- representing the variation in the architectural feature not explained by size -- in the final model. The resulting mixed-effects model has the following form:

\begin{equation}
\log(E_{i,j}) = \beta_0 + \beta_1 \log(P_i) 
+ \boldsymbol{\gamma}^\top \mathbf{R}_i
+ u_j + \epsilon_{i,j}
\end{equation}

\noindent
where
\begin{align*}
E_{i,j} & \text{ is the energy per token of model $i$ on GPU $j$,} \\
\beta_0 & \text{ is the baseline log-energy per token,} \\
\beta_1 & \text{ captures the Empirical Energy Scaling Law,} \\
\mathbf{R}_i &= \begin{bmatrix} L_i^{\perp} & H_i^{\perp} & A_i^{\perp} & K_i^{\perp} & T_i \end{bmatrix},\text{are residualized} \\
& \text{ architectural features orthogonal to } \log(P_i), \\
T_i & \text{ is the model type (categorical),} \\
u_j & \text{ is the random intercept for GPU $j$,} \\
\epsilon_{i,j} & \text{ is the residual error.}
\end{align*}

In addition, numerical architectural features are standardized to zero mean and unit variance before model fitting, to allow the fixed-effect coefficients to be interpreted as relative importance for energy efficiency.

The results confirm that size is the greatest contributor, with coefficient $0.282$ ($p < 0.001$), and show that several architectural characteristics significantly affect energy consumption beyond size:

\begin{itemize}
    \item \textbf{Key-value head count} has the strongest independent association with higher energy ($\beta = 0.121$, $p < 0.001$), so configurations with more key-value heads are systematically more energy-intensive.
    \item \textbf{Attention head count} also increases energy consumption ($\beta = 0.070$, $p = 0.004$).
    \item \textbf{Hidden size} shows a negative conditional effect ($\beta = -0.098$, $p = 0.026$), suggesting that, for a fixed parameter count, models that are wider may be relatively more energy-efficient, potentially reflecting improved hardware utilization.
    \item \textbf{Number of layers} shows a weaker, marginally significant positive association ($\beta = 0.028$, $p = 0.062$).
\end{itemize}

Finally, the estimated random-effect variance (Group Var = 0.072, reported in Table~\ref{tab:mixed_model_B}) confirms that GPU-specific factors -- such as GPU architecture, memory bandwidth, TFLOPs -- introduce non-negligible baseline differences, justifying the use of mixed-effects modeling to account for GPU-specific baselines. 

\textbf{Answering RQ2}, our analysis reveals two key insights. First, \textbf{model size is the dominant factor that influences} energy consumption, following a sublinear scaling relationship (exponents 0.237 and 0.282, both with $p < 0.001$). Second, \textbf{model architecture design introduces deviations} around this baseline. Key-value and attention head counts increase energy consumption while wider hidden dimensions reduce it, showing that parameter allocation across the model architecture materially influences inference energy efficiency.

This reinforces our claim that LLM inference energy efficiency is fundamentally a systems-level problem: both model architectural properties and GPU specifications contribute to energy cost. Consequently, optimizing inference efficiency cannot rely on model design alone, but requires software–hardware co-design and hardware-aware model placement to achieve meaningful energy savings.

\subsection{RQ3: Deployment Scenario Sensitivity}
\textit{How does the deployment scenario (batch vs. server load) influence the energy efficiency of LLMs on a given GPU? Are the same LLM–GPU pairs optimal across scenarios?}

%LLMs are used in a wide range of tasks and scenarios. 
So far, our analysis has focused on batch scenarios. However, LLMs are often used in server deployment scenarios, where resources are continuously awaiting incoming requests, producing distinct energy consumption patterns.
Server scenarios introduce idle periods driven by the stochastic arrival of requests. In our data, mean idle time rises from 2\% in the batch scenario to 53\% (high load) and 96\% (low load) in server scenarios. Consequently, as noted in Section~\ref{subsubsec:energy-metrics}, we evaluate energy efficiency 
using mean power draw, as it captures consumption in active and idle periods.

\begin{figure}[t]
    \centering
\includegraphics[width=1.0\linewidth]{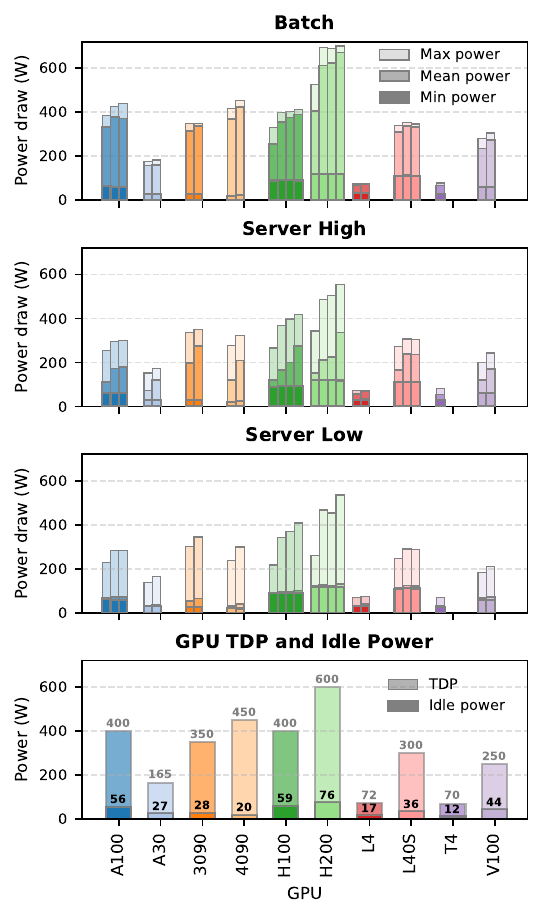}
\caption{GPU power draw across model sizes and scenarios.
Bars show the minimum, mean, and maximum power draw for different model size categories in order (S/M/L/XL) across GPUs under batch, server low, and server high load scenarios. TDP and measured idle power are reported in the last row for reference.}
\label{fig:power_draw}
\rmspace
\end{figure}

Figure~\ref{fig:power_draw} shows power draw across GPUs, model sizes and scenarios, including TDP and
idle power. TDP values are sourced from manufacturer documentation, while idle power is estimated as the mean power draw recorded by NVML prior to each inference experiment, averaged across all runs for each GPU. Mean power draw approaches the maximum in the batch scenario, while converging to idle power in server high- and low-load, especially for smaller models. This pattern highlights the importance of idle power draw in underutilized systems, which unfortunately is rarely disclosed by manufacturers, reporting TDP instead. As Figure~\ref{fig:power_draw} shows, while TDP is a useful proxy for mean power draw comparisons in batch scenarios, it is a poor proxy in server scenarios. %However, idle power draw is usually not reported by manufacturer, reporting instead TDP; which is not directly correlated to idle power as shown in the figure and is therefore not a good proxy for power draw in server scenarios.

Table~\ref{tab:gpu_ranks_server} shows the energy efficiency rankings across GPUs in the server scenario, based on mean power draw. High load rankings favor the L4 (lowest mean power draw for 19 models) and the T4 (14); low load favors the T4 (14 models) and the A30 (11); in both cases the H100 ranks third. Idle and maximum power drive these results: the L4, T4, and A30 have very low idle and max power draws (L4: 17W, 73W, T4: 12W, 82W, A30: 27W, 171W). The H100 with a relatively high idle power (59W) and mid-range max power draw (416W) benefits from its throughput advantage -- processing requests quickly reduces active time and therefore mean power draw -- outperforming other GPUs such as the L40S with a lower max power draw (357W) but lower throughput. This shows that throughput still influences energy efficiency in server scenarios, even though its effect is secondary to idle and max power.

\begin{table}[t]
\centering
\caption{GPU ranking positions based on energy efficiency under server inference.}
\label{tab:gpu_ranks_server}
\begin{tabular}{llll}
\toprule
 GPU                   & 1st   & 2nd   & 3rd   \\
\midrule
 \textbf{Server High Load}  &       &       &       \\
 \midrule
 A100 SXM4 40 GB       & 6     & 1     & 7     \\
 A30 PCIe              & 1     & 19    & 14    \\
 GeForce RTX 4090      & 0     & 0     & 1     \\
 H100 NVL 94 GB        & 9     & 6     & 4     \\
 H200 NVL              & 0     & 7     & 5     \\
 L4                    & 19    & 13    & 1     \\
 L40S                  & 0     & 1     & 1     \\
 Tesla T4              & 14    & 1     & 1     \\
 Tesla V100 & 1     & 2     & 9     \\
 \midrule
 \textbf{Server Low Load}   &       &       &       \\
 \midrule
 A100 SXM4 40 GB       & 8     & 0     & 0     \\
 A30 PCIe              & 11    & 12    & 9     \\
 GeForce RTX 3090      & 0     & 0     & 1     \\
 GeForce RTX 4090      & 5     & 12    & 15    \\
 H100 NVL 94 GB        & 8     & 5     & 3     \\
 H200 NVL              & 0     & 8     & 1     \\
 L4                    & 4     & 10    & 8     \\
 L40S                  & 0     & 0     & 5     \\
 Tesla T4              & 14    & 0     & 1     \\
 Tesla V100 & 0     & 3     & 0     \\
\bottomrule
\end{tabular}
\end{table}

\begin{figure}[t]
    \centering
\includegraphics[width=1.0\linewidth]{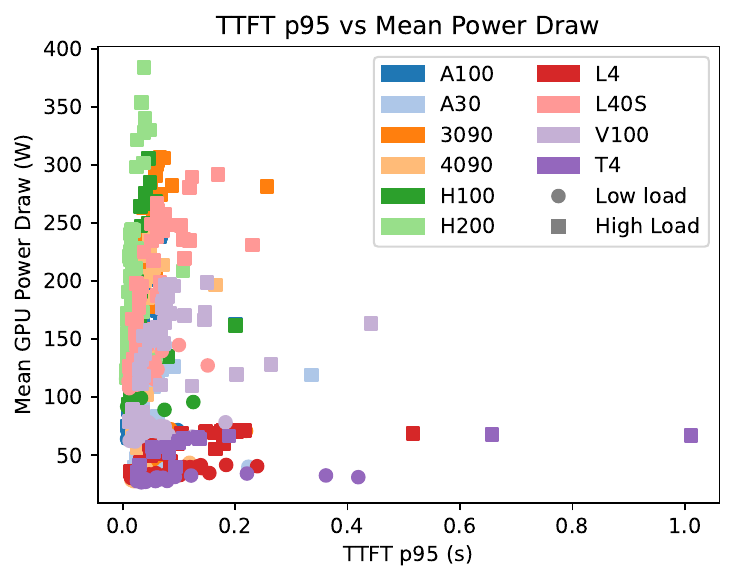}
\caption{Scatter plot showing mean GPU power draw versus the 95th percentile of Time-to-First-Token (TTFT), per model and scenario. Each data point is a (model, GPU, scenario) configuration, averaged over five iterations.}
\label{fig:power_draw_ttft}
\rmspace
\end{figure}

In real-world settings, response time requirements are critical. 
Figure~\ref{fig:power_draw_ttft} plots mean power draw versus the 95th percentile of Time-to-First-Token (TTFT) as measured by vLLM, defined as the time from the arrival of the request until the first token is generated (equal to the sum of queue and prefill time), for all models in Watt Counts, a common metric for evaluating response time in server scenarios. The T4 and L4 GPUs achieve the lowest power draw for most models but their high TTFTs limit their use in real-word applications. Conversely, the H200 consistently achieves the lowest TTFTs but with the highest mean power draw. In addition, high load scenarios require more energy and exhibit higher TTFTs than low load scenarios across all GPUs. 
A practical approach for energy efficiency is to define a response time requirement and select the GPU with the lowest mean power draw that satisfies it, given the model and expected request load. For illustration purposes, we could define this TTFT threshold of $70ms$ based on human-computer interaction research, which identifies $69ms$ as the upper limit for a system response to be perceived as instantaneous by a user after tapping a button~\cite{deber2015much}. 

Tables~\ref{tab:efficient_latency_power_server_low} and \ref{tab:efficient_latency_power_server_high} exhibit the mean power and 95th percentile of TTFT across model size categories and GPUs for the low load and high load server scenarios, respectively. By selecting the GPU with the lowest mean power draw and response time requirement (TTFT$<70ms$), it is possible to save significant energy depending on the model size and scenario. For instance, for deploying a medium-sized model such as Llama 3.1 8B, it is possible to save up to 70\% of energy by using an A30 compared to the H200, the highest-throughput GPU in our study, or 60\% when compared to the current industry standard, the H100 in the low load scenario. Another interesting case emerges for high load scenarios and large models, where the older-generation A100 achieves 10\% energy savings over the newer H100, suggesting that newer GPU generations do not always translate to better energy efficiency across all deployment scenarios. Our results show that energy-optimal GPU selection is highly deployment-specific, offering system operators actionable trade-offs between performance and energy.

\begin{table}[t]
\centering
\caption{Tail latency (TTFT p95) per experiment, averaged per model, then averaged per size category. Scenario: Server Low Load. The GPU with the lowest mean power per category that satisfies the requirement of a TTFT$<70ms$ is in bold text.}
\label{tab:efficient_latency_power_server_low}
\begin{tabular}{lll}
\toprule
 GPU                   & TTFT p95 (ms)   & Mean Power (W)   \\
\midrule
 \textbf{Small}        &                &                  \\
 \midrule
 A100 SXM4 40 GB       & 12          & 67.4             \\
 A30 PCIe              & 33          & 32.3             \\
 GeForce RTX 3090      & 57          & 52.3             \\
 \textbf{GeForce RTX 4090}      & \textbf{21}          & \textbf{31.9}             \\
 H100 NVL 94 GB        & 17          & 90.8             \\
 H200 NVL              & 08          & 120.3            \\
 L4                    & 54          & 33.1             \\
 L40S                  & 17          & 113.6            \\
 Tesla T4              & 95          & 29.6             \\
 Tesla V100 & 28          & 66.2             \\
 \midrule
 \textbf{Medium}       &                &                  \\
 \midrule
 A100 SXM4 40 GB       & 26          & 70.7             \\
 \textbf{A30 PCIe}              & \textbf{60}          & \textbf{38.3}             \\
 GeForce RTX 3090      & 88          & 65.2             \\
 GeForce RTX 4090      & 39          & 41.1             \\
 H100 NVL 94 GB        & 29          & 94.7             \\
 H200 NVL              & 13          & 125.1            \\
 L4                    & 132          & 38.8             \\
 L40S                  & 42          & 123.5            \\
 Tesla V100 & 49          & 72.8             \\
 \midrule
 \textbf{Large}        &                &                  \\
 \midrule
 \textbf{A100 SXM4 40 GB}       & \textbf{31}          & \textbf{71.1}             \\
 H100 NVL 94 GB        & 24          & 96.6             \\
 H200 NVL              & 13          & 124.7            \\
 L40S                  & 53          & 123.1            \\
 \midrule
 \textbf{Xlarge}       &                &                  \\
 \midrule
 \textbf{H100 NVL 94 GB}        & \textbf{27}          & \textbf{101.7}            \\
 H200 NVL              & 21          & 131.5            \\
\bottomrule
\end{tabular}
\end{table}

\begin{table}[t]
\centering
\caption{Tail latency (TTFT p95) per experiment, averaged per model, then averaged per size category. Scenario: Server high load. The GPU with the lowest mean power per category that satisfies the requirement of a TTFT$<70ms$ is in bold text.}
\label{tab:efficient_latency_power_server_high}
\begin{tabular}{lcc}
\toprule
 GPU                   & TTFT p95 (ms)   & Mean Power (W)   \\
\midrule
 \textbf{Small}        &                &                  \\
 \midrule
 A100 SXM4 40 GB       & 17          & 112.7            \\
 A30 PCIe              & 31          & 72.5             \\
 GeForce RTX 3090      & 25          & 198.0            \\
 GeForce RTX 4090      & 23          & 121.1            \\
 H100 NVL 94 GB        & 18          & 119.9            \\
 H200 NVL              & 09          & 151.7            \\
 \textbf{L4}                    & \textbf{67}          & \textbf{55.7}             \\
 L40S                  & 25          & 166.6            \\
 Tesla T4              & 125         & 54.2             \\
 Tesla V100            & 41          & 119.9            \\
 \midrule
 \textbf{Medium}       &                &                  \\
 \midrule
 A100 SXM4 40 GB       & 48          & 172.9            \\
 A30 PCIe              & 81          & 121.5            \\
 GeForce RTX 3090      & 73          & 275.8            \\
 GeForce RTX 4090      & 56          & 208.8            \\
 \textbf{H100 NVL 94 GB}        & \textbf{39}          & \textbf{165.9}            \\
 H200 NVL              & 21          & 212.3            \\
 L4                    & 203          & 68.7             \\
 L40S                  & 74          & 237.6            \\
 Tesla V100 & 100          & 171.3            \\
 \midrule
 \textbf{Large}        &                &                  \\
 \midrule
 \textbf{A100 SXM4 40 GB}       & \textbf{59}          & \textbf{179.0}            \\
 H100 NVL 94 GB        & 29          & 200.5            \\
 H200 NVL              & 22          & 226.0            \\
 L40S                  & 103          & 236.4            \\
 \midrule
 \textbf{Xlarge}       &                &                  \\
 \midrule
 \textbf{H100 NVL 94 GB}        & \textbf{49}          & \textbf{275.3}            \\
 H200 NVL              & 40          & 336.7            \\
\bottomrule
\end{tabular}
\end{table}

\textbf{Answering RQ3:} \emph{the deployment scenario strongly influences the energy efficiency and GPU rankings. In the server scenario, idle and max power of the GPU are main considerations while throughput becomes secondary for energy consumption, though it remains critical for meeting response time requirements.} Moreover, TDP is a poor proxy for power draw in server scenarios and idle power, an essential characteristic for assessing the energy efficiency of GPUs in this scenario, is rarely reported by manufacturers.
\section{Discussion \& Limitations}
\label{sec:discussion}
\textbf{Single-GPU scope.} Our current work is limited to LLMs deployed in single-GPU configurations, common in edge, on-prem, and cost- or energy-constrained inference systems. While this scope allows us to thoroughly characterize GPU-level power dynamics across heterogeneous architectures in isolation -- avoiding confounding effects of communication overhead, tensor-parallel synchronization, and interconnect bandwidth variability -- future work should extend this analysis to multi-GPU deployments, where additional system-level effects may alter the energy efficiency rankings observed here. Nonetheless, the single-GPU insights are directly applicable to the growing class of edge-cloud and resource-constrained deployments.

\textbf{Precision and quantization.} All evaluated models use 16-bit weights to control the dimensionality of the experimental space. Quantized models (e.g., INT8, INT4) are increasingly common in practical deployments and are known to alleviate total memory usage, memory bandwidth bottlenecks, and improve throughput, which could shift the energy efficiency ranks observed here, in particular for GPUs where memory bandwidth is their primary constraint. However, we expect the qualitative insights to generalize (e.g., total energy being the result of both GPU and model properties, the importance of memory bandwidth, the effect of model architecture properties on energy consumption, the role of idle power in server scenarios), the specific GPU rankings and model architectural properties coefficients may differ for quantized workloads. Extending Watt Counts to quantized models is a natural and important direction for future work, and the open benchmark is designed to support exactly this kind of community contribution.

\textbf{Inference engine.} We adopt vLLM, a state-of-the-art, widely adopted inference engine that has been shown to achieve leading efficiency in scalable settings~\cite{park2025survey, stuhlmann2025bench360, fernandez2025energy}. This choice ensures that observed energy–performance trade-offs primarily reflect architectural and workload characteristics rather than implementation inefficiencies. However, results may differ with other engines such as TensorRT-LLM or SGLang, particularly for specific GPU-model combinations where engine-specific optimizations may have outsized effects.

\textbf{Energy versus accuracy trade-offs.} Task accuracy and fairness are out of scope in this work, as our goal is comparative characterization of energy consumption across hardware and deployment scenarios rather than model evaluation. However, we recognize that optimizing deployments solely for energy may introduce trade-offs with output quality. These considerations should be incorporated in future multi-objective deployment frameworks that jointly optimize energy, latency, and accuracy.

\textbf{Host consumption and cooling.} Our measurements focus on GPU energy consumption, which we validate as the dominant component (78.7–92.5\%) and the one with the highest Pearson correlation against the total energy consumed across scenarios (0.995-1.0) . Total host consumption, including CPU, DRAM, storage, and cooling overhead, is not captured. In real-world data center deployments, cooling can add 30–50\% to total energy consumption depending on the power usage effectiveness (PUE) of the facility, and should be accounted for in full system-level carbon analyses.

\textbf{Carbon footprint of this work.} The dataset creation and experiments were run in data centers powered from the grid in Switzerland, Spain, and Czechia, consuming approximately 180 kWh with estimated emissions of 17 kgCO2. In particular, most experiments were run on servers sourced with a 99\% share of renewable energies. We believe that the potential energy savings enabled by reducing redundant experimentation across the community, and by enabling more informed hardware selection and deployment decisions, significantly outweigh this one-time footprint.
\section{Conclusion}
\label{sec:conclusion}
This work presents Watt Counts, a large-scale, open, energy-aware dataset and an empirical study of energy efficiency in LLM inference across 50 models and 10 heterogeneous GPUs in batch and server scenarios. Our benchmark and dataset provide foundational resources for hardware-aware deployment decisions in heterogeneous LLM inference systems, while our findings yield insights for hardware-aware placement and more general energy-efficient, sustainable LLM inference.

First, \textbf{GPU energy efficiency rankings are not universal -- they vary significantly with model size and deployment scenario}. The H100 dominates in batch scenarios for most models, but low power GPUs such as the L4 and A30 outperform it for small models, and server scenarios alter rankings entirely. This highlights the importance of hardware-aware model placement. Second, \textbf{while model size is the primary factor in energy consumption, architectural properties such as depth, hidden size, and key–value head configuration have significant effects}. As a consequence, models with similar sizes significantly vary in their energy consumption. Third, \textbf{we show that TFLOPS per Watt is a poor proxy for energy efficiency at the system level across full inference workloads}, as one of the main bottlenecks in LLM inference is memory bandwidth. 

Fourth, within our evaluation scope, \textbf{the dominant factors driving energy efficiency differ fundamentally between batch (offline) and server (online) deployments}. In batch scenarios, where energy is attributed to a finite workload, throughput and TDP are the main factors. In server scenarios, where the system runs continuously, idle and max power dominate. However, these are rarely disclosed by manufacturers, who report TDP instead. This has direct implications for hardware selection in production LLM deployments. Finally, \textbf{energy efficiency must be jointly considered with service-level requirements}, such as response time, as GPUs with low power draw might suffer from high response times, limiting their use in practical applications, underscoring the need for multi-objective evaluation.

We release the Watt Counts dataset to support future work on heterogeneous  deployments, scheduling, and benchmarking for scalable and sustainable LLM inference.

\printcredits

\section{Declaration of competing interest}
The authors declare that they have no known competing financial interests or personal relationships that could have appeared to influence the work reported in this paper.

\section{Use of Generative AI}
During the preparation of this manuscript, the authors used generative AI tools, including Claude (Anthropic) and ChatGPT (OpenAI), to assist with language refinement and clarity. All AI-generated suggestions were carefully reviewed and edited by the authors, who take full responsibility for the accuracy, integrity, and originality of the published work.

\section{Research data availability}
The Watt Counts dataset and benchmark will be publicly released on Github with an MIT License upon acceptance and are submitted as supplementary material.

%\bibliographystyle{cas-model2-names}
\bibliographystyle{elsarticle-num}

\bibliography{cas-refs}

\appendix
\section{Full List of Evaluated Large Language Models}
We list the 50 models evaluated in our experiments in Table~\ref{tab:model_list}, classifying them by their model category as discussed in Section~\ref{subsec:rq1}.

\section{Validation of GPU as the Primary Energy Consumer}
We utilize EnergyMeter to measure the energy consumption of CPU and memory DRAM with Intel RAPL and the energy consumption of the GPU with NVML on a bare metal instance (Intel(R) Core(TM) i9-9900K @ 3.60GHz, 32 GB RAM, a single NVIDIA GeForce RTX 4090), calculating the share of energy represented by the CPU, DRAM, and GPU over the total as well as measuring their empirical correlation in our experiments. Table~\ref{tab:energy_breakdown} shows the results. GPU consumes 92.5-78.7\% of the total energy consumption across all scenarios, with the highest correlation in all cases.

\begin{table}[ht]
\caption{Mean energy share and Pearson correlation with total system energy consumption per component and deployment scenario in the bare metal instance with RAPL access.}
\label{tab:energy_breakdown}
\centering
\begin{tabular}{llrr}
\toprule
\textbf{Scenario} & \textbf{Component} & \textbf{Share (\%)} & \textbf{Corr. with Total} \\
\midrule
Batch & GPU & 92.5 & 1.000 \\
 & CPU & 7.0 & 0.989 \\
 & DRAM & 0.5 & 0.989 \\
\midrule
Server high & GPU & 91.2 & 1.000 \\
 & CPU & 7.4 & 0.995 \\
 & DRAM & 1.4 & 0.127 \\
\midrule
Server Low & GPU & 78.7 & 0.995 \\
 & CPU & 16.6 & 0.575 \\
 & DRAM & 4.6 & 0.351 \\
\bottomrule
\end{tabular}
\end{table}

\section{Outlier Analysis: Energy per Token vs. Model Size}
This section details the outliers identified in the log-log regression of energy per token against model size discussed in Section~\ref{subsec:rq2}. Outliers exhibit clear structural patterns and are shown in Table~\ref{tab:outliers}.
Most outliers come from models running on the Tesla V100 and Tesla T4, older GPUs and possess limited memory bandwidth and reduced support for modern kernels leading to disproportionately high energy per token. 

In addition, two models emerge as outliers on modern GPUs apart from the V100 and T4. \texttt{SmolLM2-1.7B-Instruct} appears as an outlier on the L40S, RTX 3090, T4, and Tesla V100, consuming disproportionately high energy relative to its 1.7B parameter count. We hypothesize this is related to architecture-specific kernel inefficiencies on these devices, as the model performs within expectations on newer Ada Lovelace and Hopper GPUs. \texttt{SOLAR-10.7B-Instruct-v1.0} is another case of higher energy consumption than expected by our log-log model, appearing as an outlier on the RTX 3090, RTX 4090, and A30. This model uses a depth-upscaling architecture that duplicates specific transformer layers~\cite{kim2024solar}, increasing the depth of the original model (e.g., Mistral 7B) and its energy consumption as discussed in section~\ref{subsec:rq2}. In addition, we hypothesize the limited memory capacity of these GPUs (24GB) combined with the model size (10.7B parameters estimated to use ~21GB of RAM for its weights only), cause excessive memory pressure and degraded throughput that increases energy per token.

\begin{table}[]
    \caption{Log-log energy per token vs model size outliers}
    \label{tab:outliers}
    \centering
    \resizebox{1.0\linewidth}{!}{%
    \begin{tabular}{llrr}
\toprule
GPU & Model & Params (B) & Mean Energy/Token \\
\midrule
Tesla V100 SXM2 32 GB & allenai/OLMoE-1B-7B-0924 & 1.000000 & 0.256803 \\
Tesla T4 & microsoft/Phi-tiny-MoE-instruct & 1.100000 & 0.135182 \\
Tesla V100 SXM2 32 GB & microsoft/Phi-tiny-MoE-instruct & 1.100000 & 0.140642 \\
L40S & HuggingFaceTB/SmolLM2-1.7B-Instruct & 1.700000 & 0.113926 \\
GeForce RTX 3090 & HuggingFaceTB/SmolLM2-1.7B-Instruct & 1.700000 & 0.143895 \\
Tesla T4 & HuggingFaceTB/SmolLM2-1.7B-Instruct & 1.700000 & 0.129636 \\
Tesla V100 SXM2 32 GB & HuggingFaceTB/SmolLM2-1.7B-Instruct & 1.700000 & 0.176601 \\
Tesla T4 & microsoft/phi-2 & 2.700000 & 0.151607 \\
Tesla V100 SXM2 32 GB & ibm-granite/granite-3.1-3b-a800m-instruct & 3.000000 & 0.172620 \\
Tesla T4 & ibm-granite/granite-3.1-3b-a800m-instruct & 3.000000 & 0.190443 \\
Tesla T4 & EleutherAI/gpt-j-6b & 6.050000 & 0.305580 \\
GeForce RTX 3090 & upstage/SOLAR-10.7B-Instruct-v1.0 & 10.700000 & 0.971666 \\
GeForce RTX 4090 & upstage/SOLAR-10.7B-Instruct-v1.0 & 10.700000 & 0.490263 \\
A30 PCIe & upstage/SOLAR-10.7B-Instruct-v1.0 & 10.700000 & 0.378546 \\
Tesla V100 SXM2 32 GB & meta-llama/Llama-2-13b-chat-hf & 13.000000 & 0.630607 \\
Tesla V100 SXM2 32 GB & microsoft/Phi-3-medium-4k-instruct & 14.000000 & 0.546969 \\
Tesla V100 SXM2 32 GB & allenai/OLMo-2-1124-13B-Instruct & 14.000000 & 0.986962 \\
\bottomrule
\end{tabular}
}
\end{table}

\section{Full Mixed-Effects Model Results for LLM Properties and Energy Per Token}
In this section we include the full results for the mixed-models presented in Section~\ref{subsec:rq2}, used for understanding the effect of LLM architectural properties on energy per token. The data points used in each model correspond to the mean computed over five repeated runs for each experimental configuration (i.e., each LLM-GPU-scenario) to reduce measurement noise.

Table~\ref{tab:mixed_model_A} shows the results of mixed-effects model A, which focuses on the effect of active parameters on energy consumption and Table~\ref{tab:mixed_model_B} shows the results of model B, which includes the residuals of hidden size, number of layers, number of attention heads, and number of key value heads.

\begin{table}[t]
\centering
\small
\caption{Mixed-effects linear regression results for log energy per token (Model A: log parameters + model type).
GPU is modeled as a random intercept. Coefficients are reported with standard
errors, $z$-statistics, $p$-values, and 95\% confidence intervals.}
\label{tab:mixed_model_A}
\resizebox{1.0\linewidth}{!}{%
\begin{tabular}{lrrrrrr}
\toprule
\textbf{Variable} & \textbf{Coef.} & \textbf{Std.Err.} & \textbf{z} & \textbf{P$> |$z$|$} & \textbf{[0.025} & \textbf{0.975]} \\
\midrule
Intercept & -0.919 & 0.123 & -7.47 & 0.000 & -1.160 & -0.678 \\
log\_params & 0.237 & 0.014 & 16.93 & 0.000 & 0.210 & 0.264 \\
C(model\_type)[T.gemma2] & -0.495 & 0.109 & -4.54 & 0.000 & -0.709 & -0.281 \\
C(model\_type)[T.gpt\_oss] & -0.219 & 0.154 & -1.42 & 0.155 & -0.521 & 0.083 \\
C(model\_type)[T.granite] & -0.409 & 0.098 & -4.17 & 0.000 & -0.601 & -0.217 \\
C(model\_type)[T.granitemoe] & -0.443 & 0.105 & -4.22 & 0.000 & -0.649 & -0.237 \\
C(model\_type)[T.internlm2] & -0.377 & 0.104 & -3.63 & 0.000 & -0.581 & -0.173 \\
C(model\_type)[T.llama] & -0.383 & 0.092 & -4.16 & 0.000 & -0.563 & -0.203 \\
C(model\_type)[T.mistral] & -0.441 & 0.094 & -4.69 & 0.000 & -0.625 & -0.257 \\
C(model\_type)[T.nemotron] & -0.512 & 0.106 & -4.83 & 0.000 & -0.720 & -0.304 \\
C(model\_type)[T.nemotron\_h] & 0.174 & 0.154 & 1.13 & 0.258 & -0.128 & 0.476 \\
C(model\_type)[T.olmo2] & -0.167 & 0.103 & -1.62 & 0.106 & -0.369 & 0.035 \\
C(model\_type)[T.olmoe] & -0.082 & 0.108 & -0.76 & 0.448 & -0.294 & 0.130 \\
C(model\_type)[T.phi] & -0.221 & 0.105 & -2.10 & 0.036 & -0.427 & -0.015 \\
C(model\_type)[T.phi3] & -0.296 & 0.099 & -2.99 & 0.003 & -0.490 & -0.102 \\
C(model\_type)[T.phimoe] & -0.162 & 0.106 & -1.53 & 0.126 & -0.370 & 0.046 \\
C(model\_type)[T.qwen2] & -0.598 & 0.092 & -6.50 & 0.000 & -0.778 & -0.418 \\
C(model\_type)[T.qwen2\_moe] & -0.520 & 0.128 & -4.06 & 0.000 & -0.771 & -0.269 \\
C(model\_type)[T.qwen3] & -0.438 & 0.128 & -3.42 & 0.001 & -0.689 & -0.187 \\
C(model\_type)[T.solar] & -0.332 & 0.157 & -2.12 & 0.034 & -0.640 & -0.024 \\
\midrule
Group Var & 0.071 &  &  &  &  &  \\
\bottomrule
\end{tabular}
}
\vspace{0.5em}
\footnotesize
\textit{Model details:} 335 observations; 10 GPU groups; REML estimation;
log-likelihood = 55.81; model converged.
\end{table}

\begin{table}[t]
\centering
\small
\caption{Mixed-effects linear regression results for log energy per token (Model B: residualized architectural features).
GPU is modeled as a random intercept. Coefficients are reported with standard
errors, $z$-statistics, $p$-values, and 95\% confidence intervals.}
\label{tab:mixed_model_B}
\resizebox{1.0\linewidth}{!}{%
\begin{tabular}{lrrrrrr}
\toprule
\textbf{Variable} & \textbf{Coef.} & \textbf{Std.Err.} & \textbf{z} & \textbf{P$> |$z$|$} & \textbf{[0.025} & \textbf{0.975]} \\
\midrule
Intercept & -0.948 & 0.117 & -8.10 & 0.000 & -1.177 & -0.719 \\
log\_params & 0.282 & 0.013 & 21.69 & 0.000 & 0.257 & 0.307 \\
hidden\_size\_resid & -0.098 & 0.044 & -2.23 & 0.026 & -0.184 & -0.012 \\
num\_layers\_resid & 0.028 & 0.015 & 1.87 & 0.062 & -0.001 & 0.057 \\
num\_attention\_heads\_resid & 0.070 & 0.024 & 2.92 & 0.004 & 0.023 & 0.117 \\
num\_key\_value\_heads\_resid & 0.121 & 0.014 & 8.64 & 0.000 & 0.094 & 0.148 \\
C(model\_type)[T.gemma2] & -0.270 & 0.093 & -2.90 & 0.004 & -0.452 & -0.088 \\
C(model\_type)[T.gpt\_oss] & -0.413 & 0.177 & -2.33 & 0.020 & -0.760 & -0.066 \\
C(model\_type)[T.granite] & -0.421 & 0.092 & -4.58 & 0.000 & -0.601 & -0.241 \\
C(model\_type)[T.granitemoe] & -0.477 & 0.094 & -5.07 & 0.000 & -0.661 & -0.293 \\
C(model\_type)[T.internlm2] & -0.313 & 0.100 & -3.13 & 0.002 & -0.509 & -0.117 \\
C(model\_type)[T.llama] & -0.382 & 0.083 & -4.60 & 0.000 & -0.545 & -0.219 \\
C(model\_type)[T.mistral] & -0.375 & 0.089 & -4.21 & 0.000 & -0.549 & -0.201 \\
C(model\_type)[T.nemotron] & -0.431 & 0.092 & -4.69 & 0.000 & -0.611 & -0.251 \\
C(model\_type)[T.nemotron\_h] & 0.167 & 0.139 & 1.20 & 0.230 & -0.105 & 0.439 \\
C(model\_type)[T.olmo2] & -0.408 & 0.096 & -4.25 & 0.000 & -0.596 & -0.220 \\
C(model\_type)[T.olmoe] & 0.014 & 0.101 & 0.14 & 0.890 & -0.184 & 0.212 \\
C(model\_type)[T.phi] & -0.497 & 0.095 & -5.23 & 0.000 & -0.683 & -0.311 \\
C(model\_type)[T.phi3] & -0.419 & 0.092 & -4.55 & 0.000 & -0.599 & -0.239 \\
C(model\_type)[T.phimoe] & 0.179 & 0.150 & 1.19 & 0.234 & -0.115 & 0.473 \\
C(model\_type)[T.qwen2] & -0.448 & 0.080 & -5.60 & 0.000 & -0.605 & -0.291 \\
C(model\_type)[T.qwen2\_moe] & -0.591 & 0.132 & -4.48 & 0.000 & -0.850 & -0.332 \\
C(model\_type)[T.qwen3] & -0.397 & 0.118 & -3.36 & 0.001 & -0.628 & -0.166 \\
C(model\_type)[T.solar] & -0.427 & 0.141 & -3.03 & 0.002 & -0.703 & -0.151 \\
\midrule
Group Var & 0.072 &  &  &  &  &  \\
\bottomrule
\end{tabular}
}
\vspace{0.5em}
\footnotesize
\textit{Model details:} 335 observations; 10 GPU groups; REML estimation;
log-likelihood = 104.02; model converged.
\end{table}

\begin{table*}[t]
\caption{Full list of the 50 LLMs evaluated in this study. Models are grouped by size category 
(based on minimum GPU memory required) and sorted by total parameter count within each group. 
MoE models are listed separately as size categories are based on GPU memory fitting of dense models. 
``GPUs'' indicates the number of distinct GPUs each model was evaluated on.}
\label{tab:model_list}
\raggedright
\resizebox{0.135\linewidth}{!}{%
\begin{tabular}{llllrrr}
\toprule
\textbf{Model} & \textbf{Family} & \textbf{Arch.} & \textbf{Total Params (B)} & \textbf{Active Params (B)} & \textbf{GPUs} \\
\midrule
    \multicolumn{7}{l}{\textbf{\textbf{Small models}}}\\
    \midrule
    \texttt{openai-community/gpt2} & gpt2 & Dense & 0.124 & 0.124 & 10 \\
    \texttt{Qwen/Qwen2.5-0.5B-Instruct} & qwen2 & Dense & 0.490 & 0.490 & 10 \\
    \texttt{nvidia/AceMath-1.5B-Instruct} & qwen2 & Dense & 1.500 & 1.500 & 10 \\
    \texttt{Qwen/Qwen2.5-1.5B-Instruct} & qwen2 & Dense & 1.540 & 1.540 & 10 \\
    \texttt{HuggingFaceTB/SmolLM2-1.7B-Instruct} & llama & Dense & 1.700 & 1.700 & 10 \\
    \texttt{ibm-granite/granite-3.0-2b-instruct} & granite & Dense & 2.000 & 2.000 & 10 \\
    \texttt{google/gemma-2-2b-it} & gemma2 & Dense & 2.610 & 2.610 & 8 \\
    \texttt{microsoft/phi-2} & phi & Dense & 2.700 & 2.700 & 10 \\
    \texttt{ibm-granite/granite-3.1-3b-a800m-instruct} & granitemoe & Dense & 3.000 & 3.000 & 10 \\
    \texttt{tiiuae/Falcon3-3B-Instruct} & llama & Dense & 3.000 & 3.000 & 10 \\
    \texttt{Qwen/Qwen2.5-3B-Instruct} & qwen2 & Dense & 3.090 & 3.090 & 10 \\
    \texttt{meta-llama/Llama-3.2-3B-Instruct} & llama & Dense & 3.210 & 3.210 & 10 \\
    \texttt{microsoft/Phi-3-mini-4k-instruct} & phi3 & Dense & 3.800 & 3.800 & 10 \\
    \texttt{nvidia/Nemotron-Mini-4B-Instruct} & nemotron & Dense & 4.000 & 4.000 & 10 \\
    \texttt{nvidia/Llama-3.1-Minitron-4B-Width-Base} & llama & Dense & 5.000 & 5.000 & 10 \\
    \texttt{EleutherAI/gpt-j-6b} & gptj & Dense & 6.050 & 6.050 & 10 \\
    \midrule
    \multicolumn{7}{l}{\textbf{\textbf{Medium models}}}\\
    \midrule
    \texttt{01-ai/Yi-1.5-6B-Chat} & llama & Dense & 6.000 & 6.000 & 9 \\
    \texttt{mlabonne/NeuralBeagle14-7B} & mistral & Dense & 7.000 & 7.000 & 9 \\
    \texttt{berkeley-nest/Starling-LM-7B-alpha} & mistral & Dense & 7.000 & 7.000 & 9 \\
    \texttt{mlabonne/AlphaMonarch-7B} & mistral & Dense & 7.000 & 7.000 & 9 \\
    \texttt{nvidia/AceMath-7B-Instruct} & qwen2 & Dense & 7.000 & 7.000 & 9 \\
    \texttt{allenai/OLMo-2-1124-7B-Instruct} & olmo2 & Dense & 7.000 & 7.000 & 9 \\
    \texttt{deepseek-ai/deepseek-llm-7b-chat} & llama & Dense & 7.000 & 7.000 & 9 \\
    \texttt{internlm/internlm2-7b} & internlm2 & Dense & 7.000 & 7.000 & 9 \\
    \texttt{tiiuae/Falcon3-Mamba-7B-Instruct} & falcon\_mamba & Dense & 7.000 & 7.000 & 9 \\
    \texttt{mistralai/Mistral-7B-Instruct-v0.3} & mistral & Dense & 7.250 & 7.250 & 9 \\
    \texttt{Qwen/Qwen2.5-7B-Instruct} & qwen2 & Dense & 7.620 & 7.620 & 9 \\
    \texttt{ibm-granite/granite-3.0-8b-instruct} & granite & Dense & 8.000 & 8.000 & 9 \\
    \texttt{nvidia/Mistral-NeMo-Minitron-8B-Instruct} & mistral & Dense & 8.000 & 8.000 & 9 \\
    \texttt{meta-llama/Llama-3.1-8B-Instruct} & llama & Dense & 8.030 & 8.030 & 9 \\
    \texttt{01-ai/Yi-1.5-9B-Chat} & llama & Dense & 9.000 & 9.000 & 9 \\
    \texttt{upstage/SOLAR-10.7B-Instruct-v1.0} & llama & Dense & 10.700 & 10.700 & 8 \\
    \midrule
    \multicolumn{7}{l}{\textbf{\textbf{Large models}}}\\
    \midrule
    \texttt{google/gemma-3-12b-it} & gemma3 & Dense & 12.000 & 12.000 & 4 \\
    \texttt{meta-llama/Llama-2-13b-chat-hf} & llama & Dense & 13.000 & 13.000 & 5 \\
    \texttt{allenai/OLMo-2-1124-13B-Instruct} & olmo2 & Dense & 14.000 & 14.000 & 5 \\
    \texttt{Qwen/Qwen3-14B} & qwen3 & Dense & 14.000 & 14.000 & 4 \\
    \texttt{microsoft/Phi-3-medium-4k-instruct} & phi3 & Dense & 14.000 & 14.000 & 5 \\
    \texttt{Qwen/Qwen1.5-MoE-A2.7B-Chat} & qwen2\_moe & Dense & 14.300 & 14.300 & 4 \\
    \texttt{microsoft/phi-4} & phi3 & Dense & 14.700 & 14.700 & 4 \\
    \texttt{internlm/internlm2\_5-20b-chat} & internlm2 & Dense & 20.000 & 20.000 & 3 \\
    \midrule
    \multicolumn{7}{l}{\textbf{\textbf{XLarge models}}}\\
    \midrule
    \texttt{EleutherAI/gpt-neox-20b} & gpt\_neox & Dense & 20.000 & 20.000 & 2 \\
    \texttt{upstage/solar-pro-preview-instruct} & solar & Dense & 22.100 & 22.100 & 2 \\
    \texttt{mistralai/Mistral-Small-24B-Instruct-2501} & mistral & Dense & 24.000 & 24.000 & 2 \\
    \texttt{mistralai/Mistral-Small-Instruct-2409} & mistral & Dense & 24.000 & 24.000 & 2 \\
    \texttt{google/gemma-3-27b-it} & gemma3 & Dense & 27.000 & 27.000 & 2 \\
    \midrule
    \multicolumn{7}{l}{\textbf{\textbf{MoE models}}}\\
    \midrule
    \texttt{microsoft/Phi-tiny-MoE-instruct} & phimoe & MoE & 3.800 & 1.100 & 10 \\
    \texttt{allenai/OLMoE-1B-7B-0924} & olmoe & MoE & 7.000 & 1.000 & 9 \\
    \texttt{deepseek-ai/deepseek-moe-16b-base} & deepseek & MoE & 16.000 & 2.800 & 4 \\
    \texttt{unsloth/gpt-oss-20b-BF16} & gpt\_oss & MoE & 21.000 & 3.600 & 2 \\
    \texttt{nvidia/NVIDIA-Nemotron-3-Nano-30B-A3B-BF16} & nemotron\_h & MoE & 30.000 & 3.500 & 2 \\
\bottomrule
\end{tabular}
}
\end{table*}



\end{document}